# The style of genetic computing

Nicolas E. Buchler, Ulrich Gerland, and Terence Hwa[*]

*Physics Department, University of California at San Diego, La Jolla, CA 92093-0319*

[*]To whom correspondence should be addressed. e-mail: hwa@ucsd.edu

**ABSTRACT**

**Cells receive a wide variety of cellular and environmental signals, which must be processed combinatorially to generate specific and timely genetic responses. We present here a theoretical study on the combinatorial control and integration of transcription signals, with the finding that cis-regulatory systems with specific protein-DNA interaction and glue-like protein-protein interactions, supplemented by distal activation or repression mechanisms, have the capability to execute a wide range of control functions encoded in the regulatory DNA sequences. Using a quantitative model based on the well-characterized bacterial transcription system, we show explicitly how various regulatory logic functions can be implemented, by selecting the strengths and relative positions of the relevant protein-binding DNA sequences in the cis-regulatory region. The architecture that emerges is naturally modular and highly evolvable. Our findings suggest that the transcription regulatory apparatus is a "programmable" computing machine, belonging formally to the class of Boltzmann machines. We also expose critical shortcomings of the bacterial transcription systems, which limit the genome-wide adoption of schemes for complex transcription control, and discuss how they may be overcome by the eukaryotic transcription systems.**

Biological organisms ranging from bacteria to humans possess an enormous repertoire of genetic responses to ever-changing combinations of cellular and environmental signals. To a large extent, this repertoire is encoded in complex networks of genes closely regulating each other's activities. Characterizing and decoding the connectivity of gene regulatory networks has been an outstanding challenge of post-genome molecular biology (1-4). However, unlike integrated circuits which process information through synchronized cascades of many simple and fast nodes (millions of transistors at GHz speed), and for which connectivity is the primary source of network complexity, a gene regulatory network typically consists of only a few tens to hundreds of genes, whose expression is slow and asynchronous. Yet these "nodes" are very sophisticated in their capacity to integrate signals: Each node can be regulated combinatorially, often by 4~5 other nodes (5), and the effect of one node on another can either be activating or repressive depending on the context (6). In this study, we focus primarily on one node of a gene regulatory network, and investigate the power and limitations of combinatorial gene regulation in the context of bacterial transcription. At the end, we discuss factors limiting the genome-wide adoption of complex regulation for bacteria, and explore how they may be overcome by the eukaryotic transcription system.

**Combinatorial Transcription Control**. The activity of a gene is regulated by other genes through the concentrations of their gene products, the transcription factors (TFs). This is accomplished mechanically by the interaction of the TFs with their respective DNA targets, with each other, and with the RNA polymerase complex (RNAP) in the regulatory region of the regulated gene. Regulation can be quantified by the "response characteristics", i.e., the level of gene expression as a function of the concentrations of (activated) TFs[1]. While we consider protein concentrations as continuous variables, essential features of the response characteristics can often be represented more compactly by a binary "logic function", which specifies whether a gene is ON (expressed) or OFF (silent, or expressed at basal level) at different extremes of cellular TF concentrations, e.g., a "low" value of a few molecule per bacterium (~1 nM) or a "high" value of ~1,000 molecules per bacterium (~1 μM). In Fig. 1a, we show the logic function representations of six possible genetic responses (*g1—g6*) to two TFs, A and B. Some of these responses are commonly encountered in bacterial transcription control, e.g., the response of *g1* reflects the regulation of the well-known *lac*-operon by LacR (A) and CRP (B) in *E. coli* (7). Here, the repressive effect of A is achieved by the competitive binding of A and RNAP to the same region of DNA, and the activating effect of B is obtained from the cooperative interaction between B and RNAP when they are both bound to their sites; see Fig. 1b.

---

[1] We consider only the concentration of TFs in an *activated* state, i.e. a state which allows specific DNA binding and affects the expression of the regulated gene. Activation of TFs can be controlled by a number of mechanisms such as phosphorylation and ligand binding.

Can the same scheme be used to implement the other functions listed in Fig. 1a as well as more complex ones involving regulation by three or more TFs? Ptashne and Gann (8-9) argued that a wide range of regulatory functions can be realized by the "regulated recruitment" of TFs and the RNAP via the appropriate arrangement of their respective binding sites, provided the existence of a "glue-like attraction" between the proteins. We formulate a quantitative model based on the well-characterized bacterial transcription system to test these ideas and their extensions, and to investigate its power to implement combinatorial signal integration. The model is briefly outlined below, with details provided in *Supplementary Materials*; see also (10).

**Model.** We adopt and generalize the approach of Shea & Ackers (11), describing transcription regulation in bacteria by a thermodynamic treatment. The degree of gene transcription is quantified by the equilibrium binding probability $P$ of the RNAP to its DNA target, the promoter, given the cellular concentrations of all the activated TFs. Crucial to our model are two ingredients which we regard as the quantitative formulation of "regulated recruitment" (8,9):

(i) The binding strength of a TF-binding site on the DNA (an operator) is assumed to be *continuously tuneable* across and beyond the relevant range of cellular protein concentrations (*1~10,000* nM) through choice of the binding sequence. This feature of the protein-DNA interaction is supported by extensive experimental studies on exemplary TFs (12) and is expected on theoretical grounds for a large class of bacterial TFs (13). Evidence supporting the use of this feature by bacteria has been found in the regulation of *E. coli* flagellum synthesis (14) and SOS response (15). In our model, we quantify the TF-DNA binding strength of a site *i* by an effective dissociation constant $K_i$ which is defined as the TF concentration for half-maximal binding; see *Supp. Materials* for details. The above discussion implies that $K_i$ can be readily tuned in the range of *1~10,000* nM individually for each site *i*.

(ii) A weak, glue-like interaction of $E_{int}$ ~ *-2* kcal/mol between two proteins (TFs and/or RNAP) is possible if the relative placements of the DNA binding sites allow for direct contact of appropriate regions of the proteins. This is known for a number of well-studied proteins (9,11). On the molecular level, weak glue-like interactions can occur, for instance, via contact of hydrophobic patches of two proteins without affecting the DNA-binding function (16). Of course, it is also possible to arrange a repulsive interaction ($E_{int} = +\infty$) between two proteins by overlapping their respective binding sites, or to have no effective interaction ($E_{int} = 0$) between two proteins. The latter can be achieved by placing the binding sites for the two proteins on opposite sides of the DNA or at an appropriate distance, so that the proteins will not bind to their sites and contact each other simultaneously. Quantifying the interaction between two proteins bound to two sites *i* and *j* by a cooperativity factor $\omega_{i,j} = e^{-E_{int}/RT}$ where $RT \approx 0.6$ kcal/mol, we see that interaction between each pair of sites can be selected from the values $\omega_{i,j} = \{ 0, 1, \sim20 \}$ just by arranging the positions of the binding sites in the regulatory region.

Given the binding strengths $K_i$'s and the cooperativity factors $\omega_{i,j}$'s for all the DNA sites, the binding probability *P* of the RNA polymerase to the promoter can be straightforwardly obtained; see (10,11) and *Supp. Materials*. The task of implementing various regulatory functions is then reduced to arranging the binding sites in the cis-regulatory region such that the interaction parameters $K_i$'s and $\omega_{i,j}$'s produce the desired *P* for the various TF concentrations.

**Cis-regulatory implementations**: To illustrate how different regulatory functions can be implemented using the above model, let us consider the response of *g2* in Fig. 2a, which corresponds to the logic function **AND**, and the implementation of which is referred to as the AND-gate. It can be obtained by choosing weak binding sites for both A and B and placing them adjacent to each other (see Fig. 2a), so that each TF alone cannot bind to its site, but when both are present, binding occurs with the help of the additional cooperative interaction. This is quantitatively verified by the full response characteristics $P([A],[B])$ plotted across the range of physiological TF concentrations (1~1000 nM); see *Supp. Material* for the analytical form of $P([A],[B])$ and the interaction parameters used to achieve this response. In similar ways, one can implement the responses for the genes *g3* and *g4* (corresponding to the OR- and NAND- gate respectively); see the schematic constructs of the regulatory regions and their response characteristics in Fig. 2b and 2c, with numerical details given again in *Supp. Material*. We note that examples of these control functions can be found in natural and artificially constructed regulatory systems in bacteria (17-19), and the basic molecular mechanisms of their operations are similar to those described above.

The responses for *g5* and *g6* exemplify an increased level of complexity: The effect of an activated TF is not always activating or repressing (as is the case for *g1—g4*), but instead depends on the state of the other TF. For example, the protein B activates *g5* in the absence of protein A but represses *g5* in the presence of A, making the gene ON if either one, but not both, of the TFs are activated; this control is commonly known as the "exclusive-or" (XOR) gate. Analogous to electronic circuit design, *g5* could be achieved via a "gene cascade", e.g., by applying the gene products of *g3* and *g4* on *g2*; see Fig. 3a. More simply, the regulatory regions of *g3* and *g4* could be combined into a single region as shown in Fig. 3b, which achieves the desired characteristics without the need of any intermediate genes, thereby avoiding many potential problems associated with their expressions (e.g., time delay and stochasticity). The cis-regulatory implementation of the XOR-gate is not unique. For example, one can achieve comparable performance by using two promoters positioned sequentially in the regulatory region, with one promoter functional only when B is activated and A is not (as in Fig. 1b), and *vice versa* for the other; see Fig. 3c.

The above example illustrates a fundamental difference in the style of computation between a gene regulatory network and an electronic circuit: An electronic circuit features a "deep" architecture with many layers of cascades to take advantage of the vast number of simple but fast nodes. In contrast, a gene regulatory network cannot afford many stages of cascades due to the slowness and limited number of nodes, but can adopt a "broad" architecture integrating complex computations such as the XOR-gate into a single node. The speed constraint is especially significant in bacteria, where the time scale for gene expression is a significant portion of the total growth time under optimal conditions. Indeed, the preference for a broad but shallow network architecture has been observed recently in a large-scale analysis of the *E. coli* gene regulatory circuits (4). Of course, a limited number of gene cascades can be used if speed is not a limiting factor (e.g., in eukaryotes) and may be especially useful in situations such as cell cycle control (20) and development initiation (21) where a natural temporal order exists.

**Limitations**: There are limitations to the control functions one can implement using only the two ingredients of regulated recruitment formulated so far. This is illustrated in the response of the gene *g6*, the "equivalence" or EQ-gate. As can be seen from the truth table in Fig. 1a, a strong promoter is required here to turn the gene ON when neither of the TFs are activated, while repression is needed under multiple conditions (*i.e.*, when A is activated and B is not, and *vice versa*). It is difficult to implement both repressive conditions by the direct physical exclusion of RNAP given the small size of the promoter region; Fig. 4a illustrates the kind of unrealistic scenario involved. The situation is somewhat improved in an alternative approach involving two promoters, although multiple repressions are still needed; see Fig. 4b. This turns out to be a general problem for the implementation of more complex regulatory functions, the implementation of which will generically require multiple repression conditions. An effective strategy to overcome promoter overcrowding is the repression of the promoter *from a distance*. One way to accomplish distal repression is through DNA looping mediated by protein dimerization, a strategy adopted repeatedly in bacterial transcription regulation; see, e.g., the homodimerization of AraC in *E. coli* (22). In bacteria, DNA looping can also be facilitated by DNA bending proteins such as IHF (23).

A simple strategy to implement repression under *multiple* conditions is to employ *heterodimers*, with two subunits each recognizing a distinct DNA site while associating strongly to each other (quantified by a cooperativity factor $\omega_{i,j} \sim 100$) as shown in Fig. 5a. Long-range regulation through heterodimers has been demonstrated *in vivo* using either two regulatory proteins each fused with a recognition domain according to the "two-hybrid" approach (24) or a single regulatory protein with two distinct binding domains (25). Distal repression can be implemented here by overlapping one of the binding sites, say the target of the S-subunit, with the promoter. To control the repressive effect solely by the proteins A and B, one can set up a steady background concentration of the heterodimers and make the binding strength of the distal site weak, so that the heterodimers only bind to their respective DNA targets when recruited by the appropriate TFs placed adjacent to the distal site. Binding sites for A and B can also be placed overlapping with the distal site to turn off distal repression under desired conditions. A *cis*-regulatory construct and the corresponding response characteristics of the EQ-gate, using the distal repression scheme, is shown in Fig. 5b with multiple binding sites for the R-subunit used to enforce multiple repression conditions. Alternatively, the EQ-gate could be implemented using a distal activation scheme as shown in Fig. 5c, with the target of the S-subunit located in close vicinity of the promoter so as to recruit the RNA polymerase.

**Complex Transcription Logics**

The above schemes with distal activation and repression can be readily extended to describe combinatorial control by multiple TF species. As long as the glue-like contact interaction exists between the TFs and RNAP, it is clear that one species of TF can be substituted for another by changing the TF-specific DNA binding sequences in Figs. 1-5. (See below for a detailed discussion on possible adverse effects due to the promiscuity of glue-like interactions.) More complex regulatory functions involving 3 or more inputs (TFs) can also be implemented using the molecular apparatus we describe here, by generalizing the constructs of Fig. 5b and Fig. 5c. Fig. 6a illustrates the general architecture of the regulatory region one obtains using the distal activation scheme. Note that the emerging structure is naturally modular, in the sense that the subsequence coding for a given logical expression (as indicated above the brackets) can be moved to different positions in the regulatory region without affecting the regulatory function (5,21). Since each module can be arranged to recruit RNAP on its own, the regulatory logic function implemented by this scheme is of the form

$$\mathcal{L} = \mathcal{C}_1 \text{ \textbf{OR} } \mathcal{C}_2 \text{ \textbf{OR} } \ldots \text{ \textbf{OR} } \mathcal{C}_M \qquad (1)$$

where the binary variable $\mathcal{L}$ indicates the occupation state of the promoter, and $\mathcal{C}_m$ indicates the occupation state of the binding site $R_m$ in the $m^{th}$ module. Formally, the $\mathcal{C}_m$'s are known as logical "clauses". Intuitively, Eq. (1) corresponds to selectively "switching on" rows in a logic table (see Fig. 1a) that are OFF by default.

Within each module, the recruitment of the R-subunit to its target must be accomplished molecularly through contact with TFs bound to nearby sites. This implements the logical **AND** function, leading to the following expression for each clause

$$\mathcal{C}_m = b_{m,1} \text{ \textbf{AND} } b_{m,2} \text{ \textbf{AND} } \ldots \text{ \textbf{AND} } b_{m,n(m)}. \qquad (2)$$

Here the index $i \in \{1, \ldots n(m)\}$ labels the binding sites in the $m^{th}$ module and the binary "literals" $b_{m,i}$ express the effect (activating/repressing) of a binding site on the occupation of $R_m$. For an activating site, $b=1$ (0) if the corresponding TF concentration is high (low), while the opposite is true for a repressive site. If we represent the state of the concentration (high/low) of the TF α by a binary variable $x_\alpha$ and its inverse by $\overline{x_\alpha}$, then we have $b_{m,i} \in \{x_{\alpha(m,i)}, \overline{x_{\alpha(m,i)}}\}$ where α(m,i) denotes the identity of the TF (e.g., A, B, C, etc.) targeted by sit*e i* in module *m*.

Eqs. (1)-(2) is a special form of expressing the logic function $\mathcal{L}[x_A, x_B, x_C, ...]$ which describes the dependence of the state of gene activity $\mathcal{L}$ on the TF concentrations. It corresponds to the so-called "disjunctive normal form" (DNF) familiar in computer science, see (26). It is well known that *any* binary logic function can be expressed as a DNF, and can be reduced efficiently to a minimal (i.e., the most compact) form. This observation suggests a simple recipe to guide the construction of regulatory regions to implement a wide variety of control functions: Reduce a desired logic function to its minimal DNF and then implement each clause according to the distal activation scheme as shown in Fig. 6a. There are certainly limitations to this scheme: For instance, if a clause contains too many repressive conditions, overcrowding of binding sites within a module will limit its implementation.

From the alternative implementations of the EQ-gate in Fig. 5b and 5c, we see that it may be possible to reduce the number of repressive conditions within clauses by adopting the distal repression scheme. This scheme is obtained by overlapping the binding site S with the promoter such that each clause $\mathcal{C}_m$ can repress the promoter on its own. Consequently, gene expression occurs only if *none* of the repression clauses are satisfied. The class of logic functions implementable under distal repression are of the form

$$\mathcal{L}' = \overline{\mathcal{C}_1} \text{ AND } \overline{\mathcal{C}_2} \text{ AND } ... \text{ AND } \overline{\mathcal{C}_M}, \qquad (3)$$

where the $\overline{\mathcal{C}_m}$'s are the inverse of the clauses $\mathcal{C}_m$'s as given in (2). In contrast to the DNF form, Eq. (3) intuitively corresponds to selectively "striking out" rows in a logic table that are ON by default. With the help of DeMorgan's rule (26), we can express each $\overline{\mathcal{C}_m}$ as

$$\overline{C_m} = \overline{b}_{m,1} \text{ OR } \overline{b}_{m,2} \text{ OR } \ldots \text{ OR } \overline{b}_{m,n(m)} \qquad (4)$$

with the $\overline{b}$'s are the inverse ("NOT") of the literals $b$. The generic architecture for the *cis*-regulatory implementation of logic functions, expressed according to Eqs. (3) and (4), is shown in Fig. 6b. This is known as the "conjunctive normal form" (or CNF). As is the case with DNF, all logic functions can be reduced to a minimal CNF (26).

Putting these considerations together, we see that to implement a given logic function, one can first obtain and compare the minimal CNF and DNF, and then choose the one with a fewer number of repressive conditions within clauses. Of course, using two sets of DNA-bending heterodimers, one for distal activation and the other for distal repression, the two types of architecture could also be combined. Thus, the above theoretical considerations can guide the design of *cis*-regulatory constructs for a wide variety of complex control functions. There may, however, be a practical limit to the combinatorial approach due to the slow kinetics of assembling very large molecular complexes, if there are too many clauses or too many literals within a clause.

**Molecular Computing Machine**

The transcription machinery can be regarded as a molecular computer since it is capable of complex logic computations. Specifically, the molecular components (TFs and RNAP) satisfying the two ingredients of regulated recruitment, *i.e.*, continuously tuneable protein-DNA binding strengths and glue-like contact interaction between proteins, and further supplemented by distal activation and/or repression mechanisms, constitute a flexible toolkit, a kind of molecular Lego set, which can be assembled in different combinations to perform the desired computations. This machine is a *general-purpose* computer since its function can be "programmed" at will through choices and placements of the protein-binding DNA sequences in the regulatory region. This should be contrasted with the alternative strategy of transcription control based on dedicated, complex (e.g., allosteric) molecular interactions: In the latter, complexity of the system is derived from the complexity of proteins, while in the former, complexity is derived combinatorially from the composition of the regulatory sequences (the "software") alone, without the need of tinkering with the proteins (the "hardware").

Another way to appreciate the computational power of the transcription machinery is through analogy to a "neural network": As illustrated in Fig. 7, binding sites in a regulatory region can be viewed as "neurons" in a network, with the promoter being the output neuron, and the activated TF concentrations being the inputs to the network. The occupancy of a binding site corresponds to the state of a neuron (firing or not) and the binding strength of a site becomes the "firing threshold". Molecular interactions between the proteins play the role of "synapses" which transduce signals between the neurons. This "neural network" is distinguished by two unique features: synaptic connections are symmetric (since molecular interactions are symmetric), and some neurons in the network are "hidden" (e.g., the binding sites of the heterodimers which are not linked to the controlling inputs). As shown in *Supp. Material*, such networks are mathematically equivalent to the class of "Boltzmann machines" (27), which are known to be powerful computing machines. Thus, the transcription systems we discuss are molecular realizations of the Boltzmann machine.

A neural network can be "trained" to perform complex tasks by adjusting its synaptic strengths (27). Similarly, the regulatory system we discuss can fine tune or modify its control function by adjusting molecular interactions through a combination of the *programmable* protein-DNA and protein-protein interactions. The latter is accomplished in nature by the evolution of DNA sequences in the *cis*-regulatory region. This architecture of the regulatory system makes it very *evolvable* (9,21) since it is straightforward to modify individual DNA binding sequences (through point substitutions), adjust their positions within a regulatory region (via insertions and deletions), and move/copy them from one regulatory region to another (via duplications and recombination). In contrast, it is much more difficult to evolve the TFs themselves, since they would generally be under different constraints in different regulatory regions.

**Beyond Bacterial Transcription Control**

So far we have exploited the known characteristics of bacterial transcription system and demonstrated its immense power for the combinatorial regulation of a *single* gene. However, most known examples of bacterial transcription control are much simpler than the capabilities described. On the other hand eukaryotes, which rely heavily on the type of complex combinatorial control we discussed, use a rather different (and not well-characterized) transcription system. Might there exist some crucial limitations in the schemes of combinatorial control we described, which prevent their adoption by bacteria on a genome-wide basis?

**Promiscuity of protein interaction**: A frequent criticism of the regulated recruitment principle is the reliance on rather promiscuous, glue-like interactions between proteins. For example, if all TFs in a bacterial cell (or the nucleus of a eukaryote) can interact with each other upon contact, then the many possible unintended interactions may overwhelm the required functional interactions, making it impossible for the system to perform any regulatory functions. The frequently observed specificity of TF-TF interactions in bacteria appears to support this criticism. However, a simple estimate shows that unintended interactions are actually *not* a major concern given the weakness of the glue-like interaction and the limited total TF concentration (see *Supp. Material*). For instance, with an interaction energy of $E_{int} \sim 2$ kcal/mol and a typical bacterial genome size of $5 \cdot 10^6$, the adverse effect of promiscuous TF-TF interactions is negligible as long as the total number of TF molecules in a cell is $N \leq 10^4$. Thus, at a typical TF concentrations of *~100* molecules/cell, one species of TF can interact weakly with *~100* other species before unintended interactions become an issue at all. [Applying a similar estimate to eukaryotes, one finds that one species of TF can roughly interact with *1,000* other species before unintended interactions become significant.]

Even though the weak interaction may not be detrimental to the system, there is no reason that they will be maintained over the course of evolution if not needed functionally. Indeed, it has been estimated that the loss of protein-protein interaction is a very rapid evolutionary process (28). The isolated usage of complex combinatorial control in bacteria (see below) can thus be responsible for the *apparent* specificity of TF-TF interactions in bacteria. But as long as the weak interaction between protein pairs may be acquired rapidly by evolution (28) when functional demand arises, we may assume a generic promiscuous interaction to study the capabilities of the regulatory system.

**Inter-genic cross talk**: A major limitation of the bacterial transcription machinery becomes evident when we attempt to implement the cis-regulatory constructs of Fig. 6 at a genome-wide scale. The problem is that if every gene uses the same heterodimer pair, e.g., the subunits R and S, then they will induce "cross talk" between regulatory regions of *different* genes. For instance, the recruitment of the R-subunit to a site $R_m$ in one gene can cause the recruitment of the S-subunit to site S' of a neighboring gene. This problem is compounded by the fact that there are many more possibilities for the heterodimers to participate in the unintended cross talk than the intended distal interactions. Even though intra-genic interactions generally involve DNA looping over shorter distances than inter-genic interactions, the logarithmic dependence of DNA looping energy on distance implies that inter-genic distance must be substantially (e.g., 10 times) greater than intra-genic distance before distance can be used as an effective means to prevent cross talk. An alternative way to reduce the degree of cross talk is to introduce different heterodimer pairs for different genes; however this will require many extra genes just to code for the heterodimers. Clearly, the compact bacterial genomes can support neither the vast inter-genic separation distances nor a large number of gene-specific heterodimers. Thus, the inter-genic cross talk problem may be a key obstacle for bacteria to adopt complex combinatorial control at a genome-wide scale. However, this problem does not prevent the implementation of complex control on a few isolated genes in a bacterium.

It is important to recognize that the cross talk problem is not specific to the use of heterodimers and DNA looping. Rather, it is an unavoidable consequence of the genome-wide use of *any* distal interaction mechanism, since each regulatory region must be told which gene to regulate. It is interesting to observe that eukaryotes have developed a number of strategies to cope with the cross talk problem. For example, inter-genic distances often greatly exceed the size of genes in higher eukaryotes, making distance-based controls more feasible. Also, insulating elements have been found which limit the actions of remote regulatory regions to their designated genes (29).

Given the vast differences in the molecular mechanisms of gene regulation in prokaryotes and eukaryotes (30), what aspect of our study on combinatorial control could be applicable to eukaryotes? We argue that the qualitative aspects of our study are applicable to eukaryotes regardless of mechanisms, since our main results, e.g., the correspondence of transcription machinery to the Boltzmann machine and the implementation of CNF/DNF, are predicated only on the existence of the two key ingredients of regulated recruitment (i.e. specific protein-DNA interaction and glue-like interaction between nearby proteins) along with the possibility of distal interactions, regardless of how these ingredients are implemented molecularly. Indeed, these ingredients may occur more prevalently in eukaryotes. For example, while different TFs within a given class can interact cooperatively when placed adjacently (31), e.g., via contact of hydrophobic patches (16), an effective glue-like interaction between two unrelated TFs can also be realized through competitive DNA binding with the histone complex localized to the same region (32), without physical contact between these TFs. Short-range repression (or "quenching") can be achieved in eukaryotes without the need of overlapping binding sites (33), and distal repression can be accomplished by the recruitment of various chromatin modification agents (30,33,34) without the need of DNA looping. Thus, at the qualitative level, the very different eukaryotic transcription system together with the regulated chromatin structure presents a superior molecular platform to implement complex combinatorial control.

**Discussion and Outlook**

The current knowledge on eukaryotic gene transcription is not sufficient to warrant the construction of quantitative models of transcription regulation (3). Nevertheless, we believe our results are useful in both a qualitative and quantitative way for dissecting the combinatorial transcription control of specific systems. On a qualitative level, the simplest and most natural forms of architecture in complex regulation involving multiple modules are the CNF and DNF. The CNF-like architecture (Fig. 6b) requires repression to dominate over activation, which can be accomplished in eukaryotes through the recruitment of repressing complexes such as Tup1 in yeast (34). The DNF-like architecture (Fig. 6a) requires activation to dominate over repression, and is more natural whenever genes are repressed *by default* (e.g. through the local chromatin structure). The phenotype exhibited by DNF is "enhancer autonomy", which is observed in *Drosophila* embryonic development, e.g. expression of seven-stripe *even-skipped* is activated by five separate enhancers (21,33). On a quantitative level, our model as described in *Supp. Material* provides a concrete framework to relate knowledge of *cis*-regulatory elements to complex gene expression patterns regardless of molecular mechanisms. This is possible because our model, as a realization of the Boltzmann machine, is sufficiently generic to describe a wide range of regulatory control functions. The DNA binding strengths $K_i$ and the cooperativity factors $\omega_{i,j}$ then constitute meaningful fitting parameters to relate the verified or potential binding sites (the nodes in the Boltzmann machine) to gene expression. This approach should be particularly useful in complex cases where a given TF can act both as an activator and a repressor (6), and is hence more powerful than the class of linear models currently used (36,37) to correlate gene expression and regulatory sequence information.

A complementary direction to pursue is the engineering of complex transcription control in bacteria. While problematic at the genome-wide scale due to inter-genic cross talk, the specific schemes of combinatorial control illustrated in Fig. 6 could be implemented in bacteria for isolated genes, e.g., a few plasmids each containing a gene with a specifically designed cis-regulatory control sequence. Designer regulatory sequences could be constructed using our modeling approach as a guide, followed by the fine-tuning of interaction parameters (the $K_i$'s and $\omega_{i,j}$'s) using the protocols of directed evolution (38,39). Such constructs might be used to control gene activities *in vivo* for various bioengineering applications (40,41). While many control functions can also be implemented synthetically by a *network* of genes regulating each other, as demonstrated theoretically and experimentally in a number of studies (42-45), we believe that the combinatorial *cis*-regulatory approach is advantageous in a number of ways: Since it does not involve the iterated expression of other genes, combinatorial regulation is fast and useful in instances where timely genetic response is essential (46). Furthermore, it is neither affected by stochastic fluctuations associated with transcription and translation (47), nor by unintended post-transcriptional, post-translational, or other cellular controls exerted by the host (45). A small number of such combinatorially regulated genes linked to each other in a network through amplification and feedback could then perform very complex functions.

**Acknowledgements**: This work was initiated during the program on "Statistical Physics and Biological Information" held at the Institute for Theoretical Physics in Santa Barbara in 2001. This research is supported by the NSF through grant # 0083704 and 0211308. In addition, NEB is supported by an NSF bioinformatics fellowship and TH by a Burroughs-Wellcome functional genomics award.

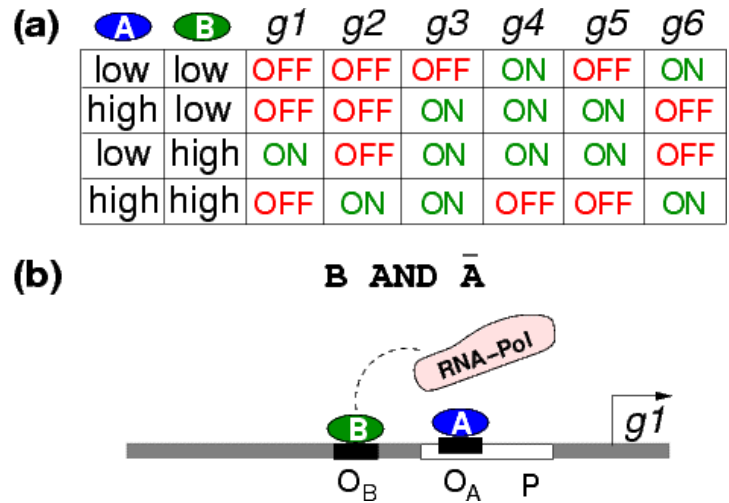

**Figure 1:** (a) Some possible gene responses (ON or OFF) according to the specific activation patterns of two transcription factors, A and B, as denoted by their cellular concentrations ("high" or "low"). (b) the *cis*-regulatory implementation for the response of gene *g1*, as adapted from the *E. coli lac*-operon. To achieve the desired effects, the operator sites need to be strong (filled boxes) and the promoter needs to be weak (open box). In this and subsequent cis-regulatory constructs, we use the offset, overlapping boxes to indicate mutual repression, and the dashed lines to indicate cooperative interaction. The logic function that this system implements is indicated above the construct, with the over line denoting the "inverse" of A, or NOT A.

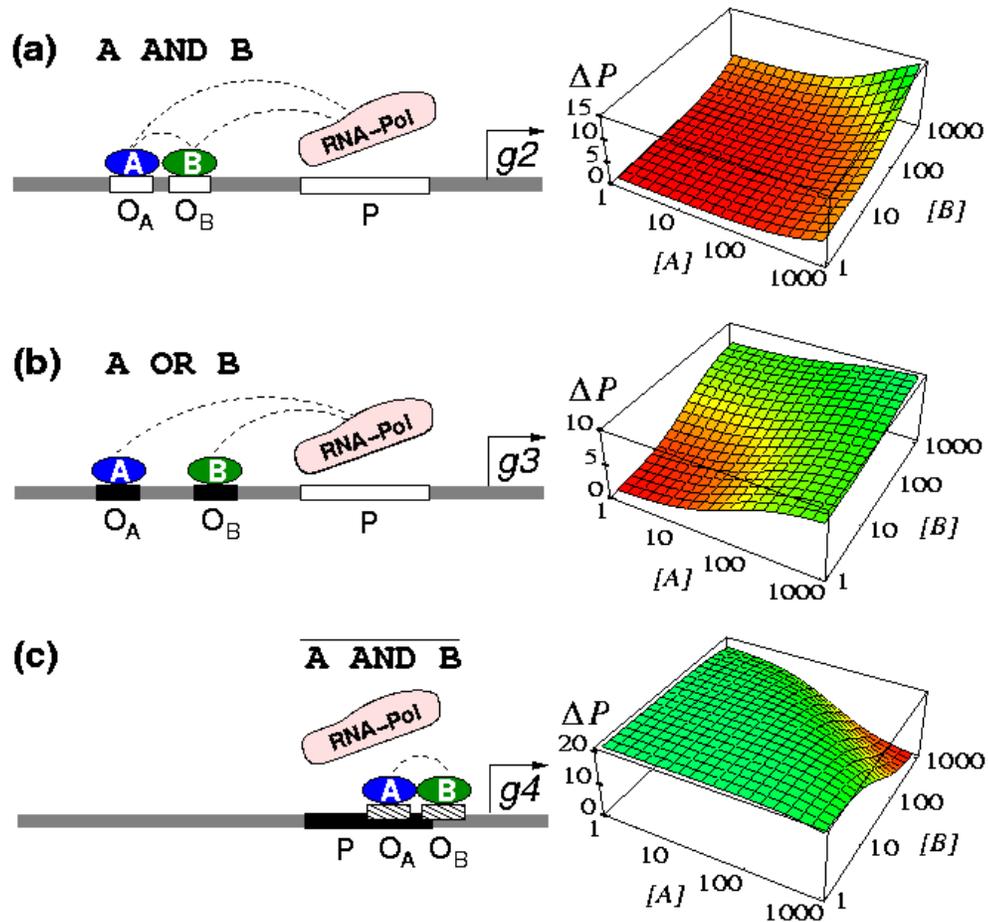

**Figure 2:** *cis*-regulatory constructs and response characteristics of the (a) AND-gate, (b) OR-gate, (c) NAND-gate. Filled, hatched, and open boxes denote strong, moderate, and weak binding sites, respectively. Dashed lines indicate cooperative interaction with $\omega_{i,j} = 20$, and overlapping boxes indicate repulsive interaction with $\omega_{i,j} = 0$. Plotted to the right of each construct is the *change* in RNAP binding probability, $\Delta P \equiv P([A],[B])/P_{min}$ for typical cellular TF concentrations [A] and [B] (in nM). The maximum fold change is $\Delta P \sim 10$ for all 3 gates. See *Supp. Materials* for the actual forms of $P([A],[B])$ and the strengths of the binding sites. Qualitative features of these plots are insensitive to the precise values of the parameters used.

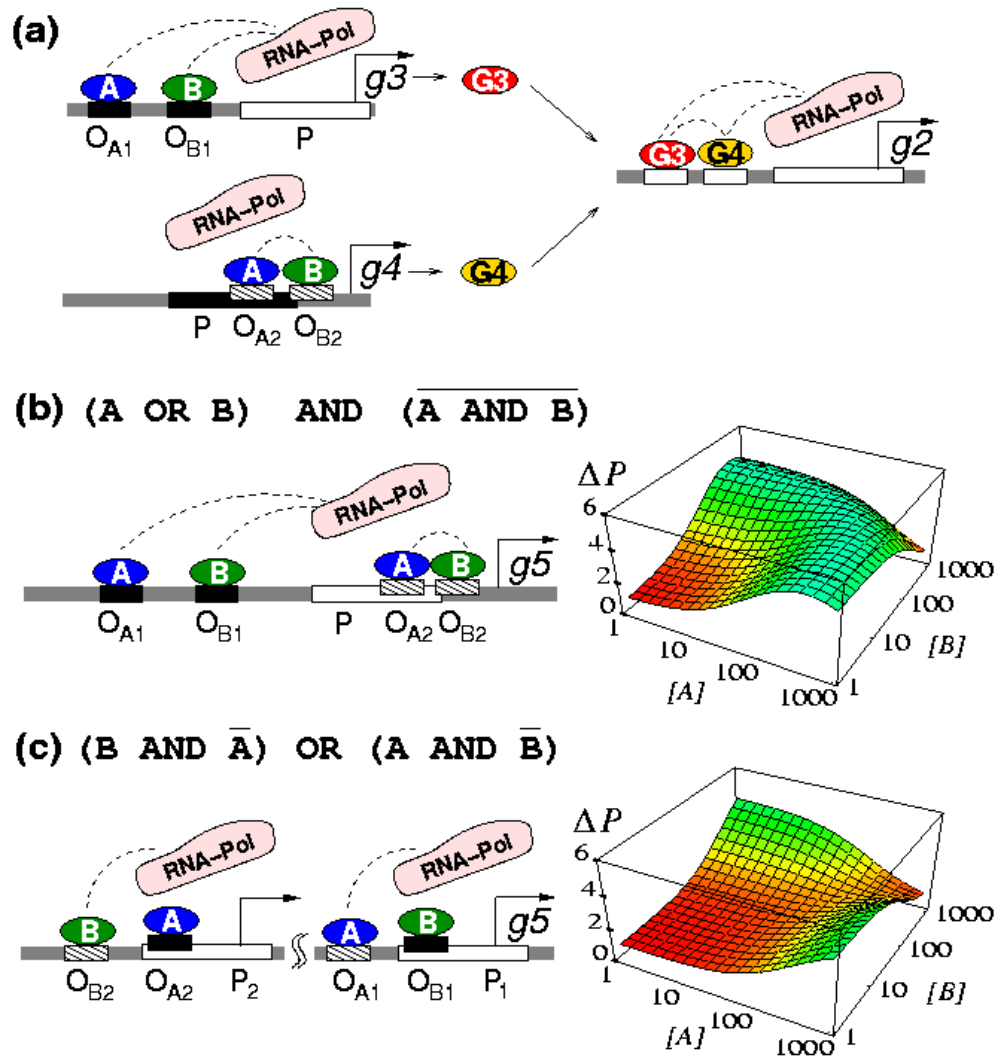

**Figure 3:** Various strategies of implementing the XOR function: (a) a gene cascade, where the intermediate gene products G3 and G4 are themselves TFs that can interact cooperatively. Alternative cis-regulatory constructs using (b) a single promoter, or (c) two promoters. Notations are the same as those used in Fig. 2. See *Supp. Materials* for the actual forms of $P([A],[B])$ and the strengths of the binding sites. The maximum fold change is ~5 in both cases.

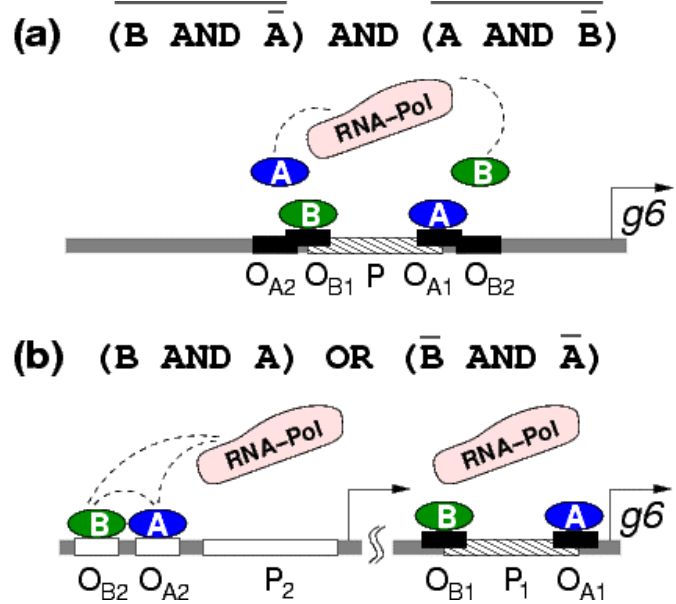

**Figure 4:** *Cis*-regulatory constructs for possible implementations of the EQ-gate, using (a) a single promoter, or (b) two promoters. Notations are the same as those used in Figs. 2-3. Both the single promoter and double promoter constructs illustrate the problem of promoter overcrowding, a situation which occurs when multiple repressive conditions are needed.

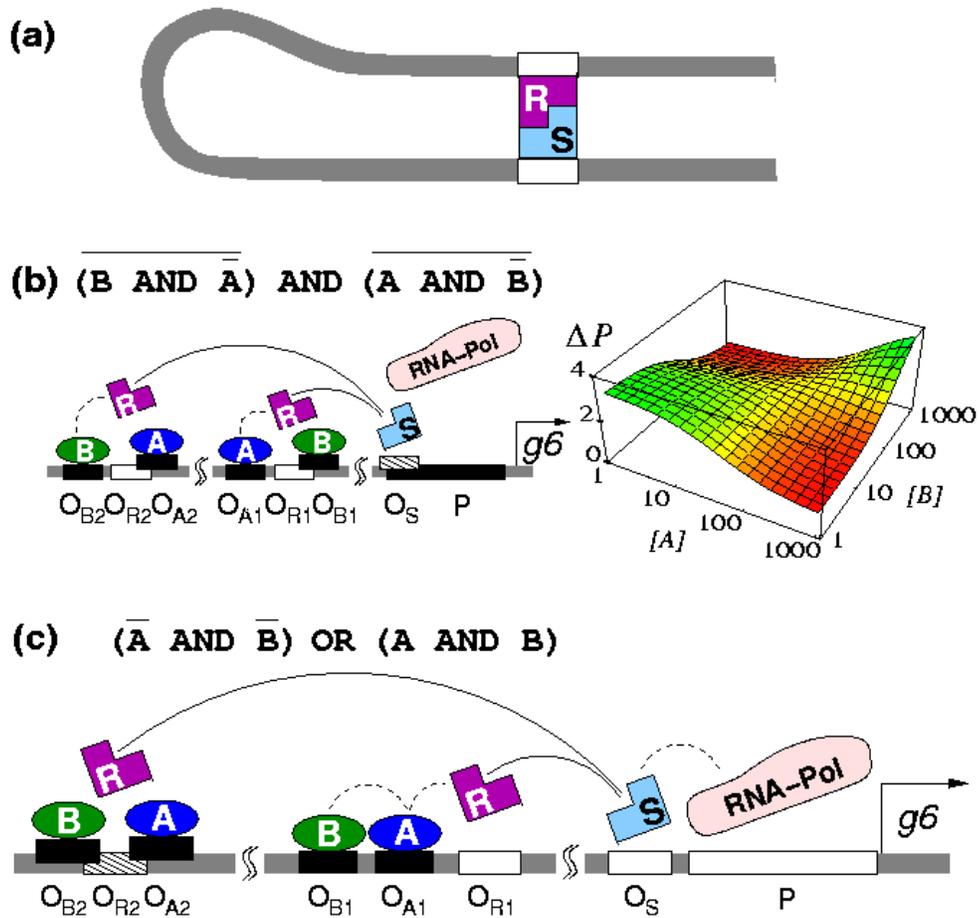

**Figure 5:** (a) Illustration of distal regulation through "DNA looping", mediated by a heterodimer formed between two subunits R and S, each recognizing a distinct DNA binding site. (b) The schematic construct of regulatory region implementing the EQ-gate (*g6* of Fig. 1a): The operators labelled R1, R2, and S are the targets of the subunits R and S as shown in (a). The solid lines indicate the relatively strong attraction between the subunits of the heterodimer. The corresponding fold-change in $\Delta P$ is shown in the response characteristics beside the construct; see *Supp. Materials* for details. (c) An alternative implementation of the EQ-gate using the distal activation mechanism.

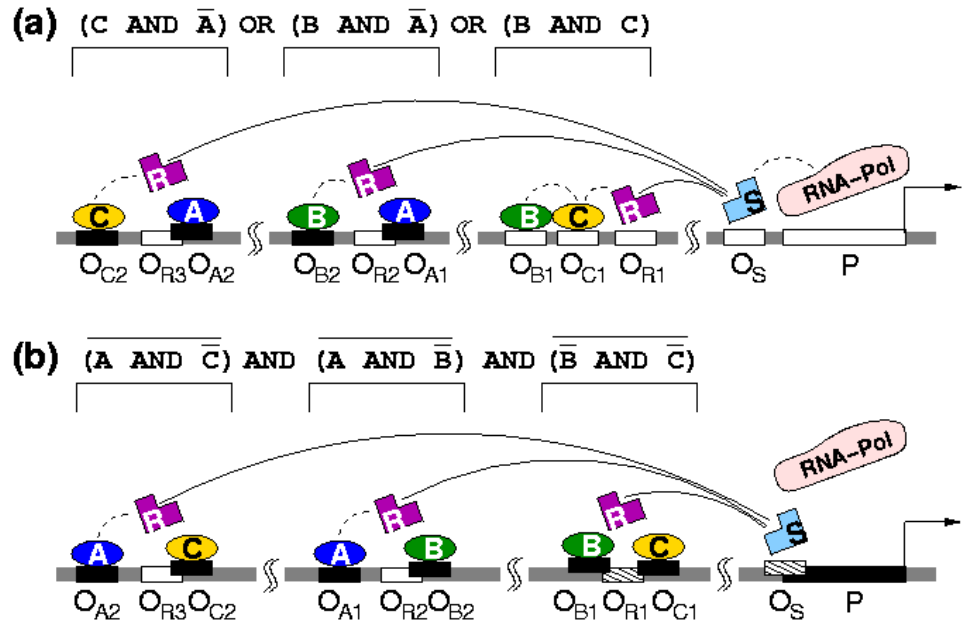

**Figure 6:** (a) Modular construct of a regulatory function involving 3 controlling TFs using the distal activation scheme. The operators labeled R1, R2, R3, and S are the targets of the recruited subunits R and the activating subunit S. Each module is bracketed with the corresponding logical syntax written above and the squiggles indicate that these modules can be at variable distances from one another. (b) The same regulatory function using the distal repression scheme.

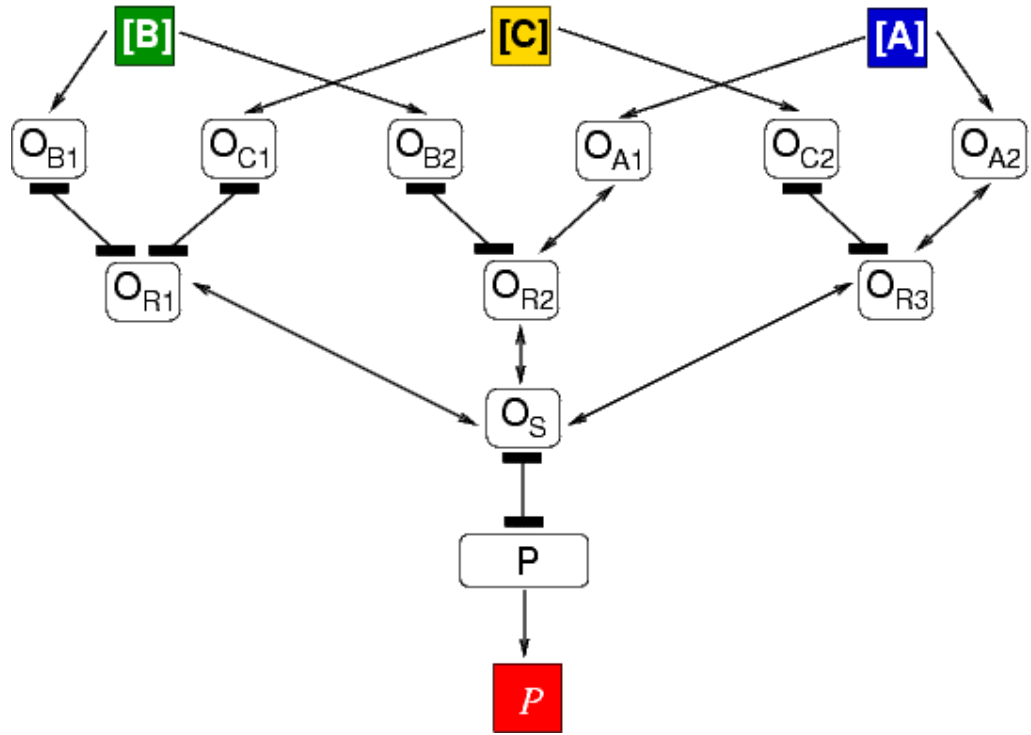

**Figure 7:** The construct of Figure 6b mapped directly on a well-studied model of neural network known as the "Boltzmann machine"; see text and *Supp. Materials*. In this mapping, the binding sites are the "neurons", the TF concentrations are the inputs, and the promoter is the output neuron. Cooperative/repressive molecular interactions between the TFs play the role of "synapses" and are denoted with arrows/bars, respectively. Note that the sites R1, R2, R3, and S are not connected to any inputs and are examples of "hidden units".

## Supplementary Material:

**Model of Transcription Regulation**

We model transcription regulation in bacteria through the thermodynamics of *pairwise* protein-DNA and protein-protein interactions (1). These interactions can be quantified by several parameters that are tuneable by the selection and placement of various protein-binding DNA sequences.

**Protein-DNA interactions**: The probability of TF-DNA binding is of the Arrhenius form (2),

$$p_i = \frac{q_i}{1+q_i} \qquad (M1)$$

where $q_i = [\text{TF}_i]/K_i$ is the binding affinity of a site $i$, $[\text{TF}_i]$ is the cellular concentration of the activated TF targeted by this site, and $K_i$ is the effective dissociation constant (relative to the genomic background) representing the concentration for half-maximal occupation of the site. It is known for a number of exemplary bacterial TFs (3) and expected on theoretical ground for a large class of bacterial TFs (4) that $K_i$ can be readily tuned across a wide range of cellular concentrations, from a low value of ~1 molecule/cell [1 nM] to a high value of ~10,000 molecules/cell [10 μM], simply by adjusting the number of bases that match the strongest binding sequence for the TF. Thus, $K_i$ is a parameter that is *individually tuneable* for each binding site. We describe RNAP-promoter binding (in the absence of any TFs) by the same Arrhenius form of Eq. (M1). Denoting the promoter as site "p", we quantify the promoter affinity by $q_p$, which is also tuneable.

**Protein-protein interaction:** Interaction between a pair of proteins (TFs or RNAP) bound to two sites $i$ and $j$ is quantified by a cooperativity factor $\omega_{i,j}$, which is tuneable to a degree by the relative placement of these sites in the regulatory region. For example, the choice $\omega_{i,j} = 0$ can be implemented by the mutual exclusion between two proteins when their binding sites are made to overlap. Interaction can also be "turned off" (described by $\omega_{i,j} = 1$) by increasing the separation between two sites (but not placing them too far), so that physical contact cannot be made when both proteins are bound to their sites.

Cooperative interaction with $\omega_{i,j} > 1$ may be obtained if two proteins can contact each other while bound to their sites. Due to its structural complexity, RNAP can contact TFs over a range of TF-binding site positions extending e.g. several tens of bases upstream of the promoter (5). This interaction is weak with typical binding free energy in the range of a few kcal/mol, corresponding to a cooperativity factor of $\omega = 10 \sim 100$. Two TFs can interact cooperatively if they are bound to adjacent sites or if they can contact each other through DNA looping. In the former case, known binding free energies are again of the order ~ *2* kcal/mol (1,9) with $\omega \sim 20$. The interaction leading to DNA looping is necessarily much stronger. It can result from two distinct proteins that bind strongly to form a heterodimer (7) as indicated in Fig. 5a or from a single protein with two fused DNA-binding domains (8). Mathematically, we model this effect by a pair of TFs each with its own DNA-binding interaction as described by Eq. (M1), along with a special cooperative interaction with cooperativity factor $\Omega \gg \omega$. Taken together, the protein-protein interaction described above can be summarized by

$$\omega_{i,j} = \{\ 0,\ 1,\ \omega,\ \Omega\ \} \tag{M2}$$

which is individually selectable for each pair of binding sites *i,j*. For simplicity, we report here only results with $\omega=20$ and $\Omega=100$. We have checked that the implementation of the logic functions in Fig. 1a are independent of the value of $\omega$ in the range *10 ~ 100*, and $\Omega \geq 100$.

**Gene transcription:** In bacteria, the rate of gene transcription is controlled in many instances by the amount of time the RNA polymerase spends bound to the promoter. Following Shea and Ackers (1), we quantify the degree of gene transcription by the equilibrium probability $P$ of RNAP-promoter binding due to interaction with bound TFs. For a single promoter, this quantity can be expressed as

$$P = \frac{Z_{ON}}{Z_{OFF} + Z_{ON}} \qquad (M3)$$

where $Z_{ON}$ and $Z_{OFF}$ are respectively the partition sum of the Boltzmann weights $W$ over all states of TF-binding for the promoter bound and not bound by the RNAP. In the simplest case involving a single TF-binding site (say site "1"), we have $Z_{OFF} = 1 + q_1$ and $Z_{ON} = q_p(1 + q_1\omega_{p,1})$. With multiple TF-binding sites (labelled as sites $i=1, ...L$), the Boltzmann weight for each configuration of site occupation is still a simple product of the $q_i$'s and $\omega_{i,j}$'s, under the assumption that the TF-TF interaction is glue-like (9,9,10). It is convenient to introduce a binary variable $\sigma_i \in \{0,1\}$ to denote the occupation of each site $i$. We have

$$W[\sigma_1,...,\sigma_L] = \prod_{i=1}^{L} q_i^{\sigma_i} \prod_{i<j} \omega_{i,j}^{\sigma_i\sigma_j} \qquad (M4)$$

as the weight for each configuration $\{\sigma_1, ..., \sigma_L\}$, with $Z_{OFF}$ obtained as the sum

$$Z_{OFF} = \sum_{\sigma_1=\{0,1\}} ... \sum_{\sigma_L=\{0,1\}} W[\sigma_1,...,\sigma_L]. \qquad (M5)$$

The expression for $Z_{ON}$ can be generally written as

$$Z_{ON} = \sum_{\sigma_1=\{0,1\}} ... \sum_{\sigma_L=\{0,1\}} Q[\sigma_1,...,\sigma_L] \cdot W[\sigma_1,...,\sigma_L] \qquad (M6)$$

where $Q[\sigma_1, ..., \sigma_L]$ is the Boltzmann weight due to the interaction of the RNAP with the bound TFs. This interaction is promoter-dependent and can be rather complicated for multiple TFs. For example, for the $\sigma^{70}$-promoters of *E. coli*, the RNAP-TF interaction can be "synergistic" (5,11,12) since two subunits of the RNAP holoenzyme can simultaneously contact two different TFs bound to upstream locations, while for the $\sigma^{54}$-promoters, the interaction is "independent" because activation of the RNAP involves binding with only one TF at a time (13). We have investigated both types of interactions and obtained similar conclusions. The response characteristics used in the text are produced by the (simpler but more restrictive) independent interaction model, given by the weight

$$Q = q_p \prod_{i=1}^{L} \left[1 - \sigma_i \delta(\omega_{0,i}, 0)\right] \cdot \left[1 + \omega \sum_{j=1}^{L} \sigma_j \delta(\omega_{p,j}, \omega)\right]. \quad (M7)$$

Here, the first bracket insures that the promoter cannot be occupied (i.e., $Q=0$) if any one of the repressor sites (those with $\omega_{p,i} = 0$) is occupied. The second bracket describes the additional weight gained by the interaction of the RNAP with all the bound TFs that it can interact cooperatively with (those with $\omega_{p,j} = \omega$), but only one at a time.

For those cases where a single gene is controlled by two promoters, we quantify the degree of gene transcription by the equilibrium probability $P$ that the RNA polymerase bind to at least one of the promoters. Assuming that there is no interaction between the two promoters (i.e., the TFs do not simultaneously interact with both polymerases in the unlikely case that both promoters are occupied), we can write the binding probability as

$$P = \frac{Z_{ON}^{(1)} \cdot Z_{OFF}^{(2)} + Z_{OFF}^{(1)} \cdot Z_{ON}^{(2)} + Z_{ON}^{(1)} \cdot Z_{ON}^{(2)}}{Z_{OFF}^{(1)} \cdot Z_{OFF}^{(2)} + Z_{ON}^{(1)} \cdot Z_{OFF}^{(2)} + Z_{OFF}^{(1)} \cdot Z_{ON}^{(2)} + Z_{ON}^{(1)} \cdot Z_{ON}^{(2)}} \quad (M8)$$

where $Z_{ON}^{(i)}$ and $Z_{OFF}^{(i)}$ are the partition sum of the Boltzmann weights $W$ over all states of TF-binding when promoter $p_i$ is bound and not bound by the RNAP, respectively.

**Implementation of Logic Gates**

Eqs. (M3)-(M8) completely specify our model of transcription regulation. To use them to compute the response of a given gene, one needs to supply the *cis*-regulatory construct specifying all the pair wise protein interactions $\omega_{i,j}$'s, as well as the affinities $q_i$'s of all the DNA sites in the regulatory region. The protein interactions can be represented graphically as in Figs. 2, 3 and 5b, with $\omega_{i,j} = 0$ if two sites overlap, $\omega_{i,j} = \omega = 20$ if two sites are linked by a dashed line, $\omega_{i,j} = \Omega = 100$ if linked by a solid line, and $\omega_{i,j} = 1$ if otherwise. The analytical expression for each of the response characteristics $P([A],[B])$ plotted in Figs. 2, 3 and 5b is then obtained by using Eqs. (M3) or (M8), as appropriate, with the corresponding expressions for $Z_{ON}$ and $Z_{OFF}$, and the values of the binding affinities as given below:

**AND-gate (Fig. 2a):**

$$Z_{OFF} = 1 + q_A + q_B + \omega \, q_A q_B$$

$$Z_{ON} = q_p (1 + \omega \, q_A + \omega \, q_B + 2\omega^2 q_A q_B)$$

$$K_A = K_B = 3{,}500; \quad q_p = 1/35$$

**OR-gate (Fig. 2b):**

$$Z_{OFF} = 1 + q_A + q_B + q_A q_B$$

$$Z_{ON} = q_p (1 + \omega \, q_A + \omega \, q_B + 2\omega \, q_A q_B)$$

$$K_A = K_B = 100; \quad q_p = 1/20$$

**NAND-gate (Fig. 2c):**

$$Z_{OFF} = 1 + q_A + q_B + \omega \, q_A q_B$$

$$Z_{ON} = q_p$$

$$K_A = K_B = 100; \quad q_p = 100$$

**XOR**-gate, single promoter (Fig 3b):

$$Z_{OFF} = (1 + q_{A2} + q_{B2} + q_{A2}q_{B2}) \cdot (1 + q_{A1} + q_{B1} + \omega\, q_{A1}q_{B1})$$

$$Z_{ON} = q_p(1 + \omega\, q_{A1} + \omega\, q_{B1} + 2\omega\, q_{A1}q_{B1})$$

$$K_{A1} = K_{B1} = 200; \quad K_{A2} = K_{B2} = 900; \quad q_p = 1/10$$

**XOR**-gate, double promoter (Fig 3c):

$$Z_{OFF}^{(1)} = (1 + q_{A1}) \cdot (1 + q_{B1})$$

$$Z_{OFF}^{(2)} = (1 + q_{A2}) \cdot (1 + q_{B2})$$

$$Z_{ON}^{(1)} = q_{p1}(1 + \omega\, q_{A1})$$

$$Z_{ON}^{(2)} = q_{p2}(1 + \omega\, q_{B2})$$

$$K_{A1} = K_{B2} = 500; \quad K_{A2} = K_{B1} = 100; \quad q_{p1} = q_{p2} = 1/20$$

**EQ**-gate, long-distance repression (Fig. 5b):

$$Z_{ON} = q_p \left(Q_{R1}^+ + Q_{R1}^-\right) \cdot \left(Q_{R2}^+ + Q_{R2}^-\right)$$

$$Z_{OFF} = \left(Q_{R1}^+ + Q_{R1}^-\right)\left(Q_{R2}^+ + Q_{R2}^-\right) + q_S \left[Q_{R1}^- Q_{R2}^- + \Omega \cdot \left(Q_{R1}^+ Q_{R2}^- + Q_{R1}^- Q_{R2}^+\right) + 2\Omega\, Q_{R1}^+ Q_{R2}^+\right]$$

$$Q_{R2}^- = (1 + q_{A2}) \cdot (1 + q_{B1}); \quad Q_{R2}^+ = q_{R2}(1 + \omega\, q_{B1})$$

$$Q_{R1}^- = (1 + q_{A1}) \cdot (1 + q_{B2}); \quad Q_{R1}^+ = q_{R1}(1 + \omega\, q_{A1})$$

$$K_{A1} = K_{B1} = 200; \quad K_{A2} = K_{B2} = 50; \quad q_{R1} = q_{R2} = 1/50; \quad q_S = 10; \quad q_p = 40$$

Note that the binding affinities $q$'s are directly selected for the promoters (site "p") and the binding sites of the auxiliary TFs R and S, since the cellular concentrations of the RNAP and the auxiliary regulators are assumed to be only weakly variable. The remaining $q$'s are defined through the variable controlling TF concentrations [$A$] and [$B$], i.e.,

$$q_A = \frac{[A]}{K_A}; \quad q_B = \frac{[B]}{K_B}; \quad q_{A1} = \frac{[A]}{K_{A1}}; \quad q_{B1} = \frac{[B]}{K_{B1}}; \quad q_{A2} = \frac{[A]}{K_{A2}}; \quad q_{B2} = \frac{[B]}{K_{B2}}$$

where $K_A$, $K_B$, $K_{A1}$, $K_{B1}$, $K_{A2}$, $K_{B2}$ are the strengths of the various sites labelled in Figs. 2, 3 and 5b. For all the logic gates implemented above, we arbitrarily considered promoter occupancy of larger than 40% as sufficient for a gene being "ON".

**Mapping to Neural Networks**

The model of transcription regulation described by Eqs. (M3)-(M8) belongs to the class of "recurrent" neural networks (14). To highlight the connection, it is convenient to recast the partition function $Z_{OFF}$ given in Eqs. (M4) and (M5) in terms of an "Ising Hamiltonian" $\mathcal{H}$, such that $Z_{OFF} = \sum_{\{\sigma\}} e^{-\mathcal{H}/RT}$. In this framework where neural network models are often described (14), we have $\mathcal{H} = \sum_{j=1}^{K} h_j \sigma_j + \sum_{i \neq j} J_{i,j} \sigma_i \sigma_j$ where each binding site $j$ is identified as a "neuron", $\sigma_j$ indicates the state of the $j^{th}$ neuron, $h_j = -RT \ln(q_j)$ is the "input" to that neuron, and $J_{i,j} = -RT \ln(\omega_{i,j})$ is the synaptic connection between the neurons $i$ and $j$. Since $h_j$ biases the neuron to the "on" state ($\sigma_j = 1$) only if $q_j > 0$ or $[TF_j] > K_j$, we can identify the binding strength $K_j$ as the "firing threshold" of the $j^{th}$ neuron. Since this network contains "hidden units", which are "neurons" not directly linked to the controlling inputs [A] and [B] (e.g., the binding sites of the auxiliary proteins R and S as shown in Fig. 7), the system is known as the "Boltzmann machine" (14)

The usual operation of neural networks (including the Boltzmann machine) amounts to finding the values of the connection matrix elements $J_{i,j}$ to implement the desired tasks, e.g., classification. The operation of the transcription control system we describe here is somewhat different: The $J_{i,j}$'s are constrained to take on one of the 4 discrete values corresponding to the form of protein-protein interaction described by Eq. (M2). Instead, it is the firing thresholds that can be tuned continuously.

**Promiscuity of protein interactions**

At high protein concentrations, a nonspecific, glue-like interaction between TFs can lead to many spurious interactions that jeopardize the intended cis-regulatory control. Here, we provide a simple estimate of the range of TF concentrations over which this problem can be safely ignored. We will only consider spurious interactions that occur while the TFs are bound to DNA, since the TF molecules spend most of the time bound to the genome (either specifically or nonspecifically) due to electrostatic attraction (15,16). For the same reason, we neglect possible spurious interactions between TFs and other, non-DNA-binding proteins. Let $N$ denote the total number of activated TF molecules in a bacterial cell at a given instant in time. The average separation distance between two such TFs along the DNA is $\ell = \Gamma/N$ where a typical genome size $\Gamma$ is $5 \times 10^6$ bp for bacteria. Two TF molecules will associate with each other if the interaction energy $E_{\text{int}}$ overcomes the entropy cost of association. The latter is approximately $RT \ln(\ell/a)$, where the microscopic length $a$ represents the range of interactions between TFs; we take *10* bp as a conservative upper bound for $a$. For a weak interaction energy of $E_{\text{int}} \sim 2$ kcal/mol, we can then safely ignore spurious interactions as long as $N \leq 10^4$. Thus, at a typical cellular concentration of *~100* molecules/cell, one species of TF can interact weakly with ~*100* other species before spurious interactions can affect cis-regulatory control at all.